\documentstyle[prd,aps]{revtex}
\begin{document}
\title{Local Spontaneous Symmetry Breaking for Film System
Within Scalar $\phi^4$ Model for Phase Transition}
\author{C.B. Yang and X. Cai}
\address{
Institute of Particle Physics, Hua-Zhong Normal University,
Wuhan 430079, China\footnote{Mailing address:yangcb@iopp.ml.org;
 yangcb@sgi31.rmki.kfki.hu}}
\date{\today}
\draft
\maketitle

\begin{abstract}
The difficulty is analysed in evaluating fluctuations in phase transition of
finite-size system at temperature far below the critical point. Film system
is discussed with one-component order parameter $\phi^4$ model for phase
transition. Non-trivial vacuum state corresponding to minimum Hamiltonian
is given approximately for various boundary conditions. It is shown that the
spontaneous symmetry breaking plays an important role for such systems, and
that perturbative calculations can be done safely when the effect of the vacuum
state or the local spontaneous symmetry breaking is taken into consideration.
\end{abstract}
\vskip 1cm

Finite-size effects near critical points have been remained over the past two
decades to be an important topic of the active research both theoretically
and experimentally [1].
Nowadays, the experimental sample are usually so pure and so well
shielded from perturbing fields that the correlation length can grow up to
several thousand angstroms as the critical point is approached.
When one or more dimensions of a bulk
system is reduced to near or below a certain characteristic length scale,
the associated properties are modified reflecting the lower dimensionality.
It is believed that finite-size effects are precursors of the critical
behavior of the infinite system and can be exploited to extract the limiting
behavior. A central role plays the finite-size scaling behavior predicted by
both the phenomenological [2] and renormalization group [3] theories. Those
theories allowed a systematic discussion of the finite-size effects and,
consequently, form the cornerstone of our current understanding of the way in
which the singularities of an infinite system are modified by the finiteness
of the system in some or all of the dimensions. Of course, the exact form of
scaling functions can't be given in those scaling theories.

In 1985, Br\'{e}zin and Zinn-Justin (BZ) [4] and Rudnick, Guo and Jasnow
(RGJ) [5] developed two field-theoretical perturbation theories for the
calculation of the finite-size scaling functions within the $\phi^4$ model
which corresponds to the Ising model. Most applications of these theories to
three-dimensional systems have been restricted to $T$ higher than the bulk
critical temperature $T_C$ [6]
with a few calculations in region below $T_C$ [7]. In recent years the $\phi^4$
and the extended $\phi^6$ models have been used to investigate the multiplicity
fluctuations in the final states for first- and second-order phase transitions
of quark gluon plasma [8, 9], under the approximation similar to the so-called
zero-mode approximation. However, some limitations exist in the theories of 
Ref. [4,5]. As pointed out in the first paper in
Ref. [10], the theory of BZ is not applicable for $T<T_C$ and the results from
RGJ theory are not quantitatively reliable in the same temperature region
since the coefficients of the Gaussian terms in the integrals are negative
for those temperatures. In Ref. [11] the order parameter
is expanded into sum of eigenfunctions of $\bigtriangledown^2$ for various
boundary conditions. Again, the functional integral is turned out into
normal integrals. But the fluctuations can be evaluated only for 
temperature not too far below the critical point.
Authors of Ref. [10] tried to avoid the difficulty
mathematically, but they failed to account  for the origin of the difficulty
physically. Although the modified perturbation method in Ref. [10]
can be used for both $T>T_C$ and $T<T_C$, the calculation is lengthy and can
be done only to the first order in practice. Since one does not know the exact
order of values of higher order terms, theoretical results have large
uncertainty.

It should be pointed out that all perturbation theories mentioned above are
based on Fourier decomposition of the order parameter. This method is
natural because the decomposition enables one to transform the functional
integral into an infinite product of tractable normal integrals. Although
such decomposition has simple physical explanation which is very fruitful
for the understanding of properties of infinite systems and can deduce
reliable physical results, as in the case of usual field theories in particle
physics, it brings about a great deal of calculations for systems with
finite-size. This is not
surprising. As is well-known, quantities complicated in coordinate space may
have simple momentum spectra thus look simpler in momentum space, but those
obviously nonzero only in a finite range must have puzzling momentum spectra.
Therefore, for the study of properties of finite-size systems, calculations
in coordinate space might be simpler and more effective. The point here
is that one must calculate the complicated functional integral which is very
difficult to be evaluated directly.

It should be asked that which physical effect causes the failure of direct
perturbative calculation of fluctuations for finite-size system with
temperature below $T_C$. In our opinion, the real origin of the difficulty
lies in the lack of knowledge about the spontaneously symmetry breaking for
finite-size systems. It is well-known that an infinite system will have
non-zero mean order parameter $\phi_0$, which is called vacuum state of
the system in this Letter since it corresponds to minimum of the Hamiltonian
$H$, if the temperature is below the critical one, and everyone knows that
the difficulty of negative coefficient for the Gaussian term can be overcome
by shifting the order parameter, $\phi\to\phi+\phi_0$. This phenomenon is
known as the spontaneous symmetry breaking because of the fact that $\phi_0$
does not have the same symmetry as $H$ does. This kind of spontaneous
symmetry breaking for infinite system can be called global since the shift
$\phi_0$ is the same for every point in the space. For finite-size
system, such a simple shift of the order parameter does not work
because of the existence of specific boundary conditions for the systems.
Anyway, fluctuations of the system, in their own sense, should be around
certain vacuum state which corresponds to minimum Hamiltonian $H$, and they
can be approximated by Gaussian terms in most cases if they are not very
large. Thus one sees that the vacuum state plays an determinative role in
the study of fluctuations in the phase transitions. For infinite system,
the vacuum state $\phi_0$ is constant and can easily be calculated.
But for finite-size systems, the vacuum is surely not constant nor it is
easy to be obtained. So, the spontaneous symmetry breaking for finite size
system can be called local one.  Therefore, the solution for the vacuum state
is non-trivial and necessary, and one has reason to hope that the difficulty
mentioned above can be overcome once the vacuum state is known.

In this Letter, we first calculate the vacuum states for $\phi^4$ model
of phase transition with one-component order parameter under various
boundary conditions. Then, with the vacuum states, the Hamiltonian of the
system is reexpressed as Gaussian term and higher order fluctuations
of a locally shifted order parameter. And it is shown that the perturbative
calculation can be done with the new Hamiltonian 
for temperatures far below the bulk critical point.

In a $\phi^4$ model for phase transition with a one-component order
parameter, the partition function can be expressed as a functional
integral of exponential of the Hamiltonian $H$ of the system   
\begin{equation}
Z=\int {\cal D}\phi \exp(-H)=\int {\cal D}\phi \exp\left\{-\int d^3\,x\left[{
\gamma\over 2}\phi^2+\frac{1}{2}(\bigtriangledown \phi)^2+\frac{u}{4!}\phi^4
\right]\right\}\ ,
\end{equation}

\noindent in which $\gamma=a(T-T_C)$, $a$ and $u$ are temperature dependent
positive constants, $\phi$ is the order parameter of the system. In the
following, we are limited only to systems of a film with thickness $L$.
Since we are interested only in the temperature region $T<T_C$ or $\gamma<0$,
the Hamiltonian $H$ can be standardized by introducing correlation length
$\xi=\sqrt{-1/\gamma}$, new order parameter $\Psi=\phi/\phi_0$ with
$\phi_0=\sqrt{-6\gamma/u}$ the vacuum state for bulk system, scaled
coordinates ${\bf r}^\prime={\bf r}/L$, and reduced thickness $l=L/\xi$, into
\begin{equation}
H=\int d^3 x^\prime {L^3\phi_0^2\over \xi^2}\left[
{1\over 2l^2}(\bigtriangledown^\prime \Psi)^2-\frac{1}{2}\Psi^2+\frac{1}{4}
\Psi^4\right]\ .
\end{equation}

\noindent From this expression one can get the equation for the vacuum state
by ${\delta H\over \delta\Psi}=0$. The vacuum state satisfies
\begin{equation}
{1\over l^2}{d^2\Psi_0\over dx^2}=-\Psi_0+\Psi_0^3\ .
\end{equation}

\noindent In the equation we have used $x$ instead of $x^\prime$ in the range
(0, 1) to denote the coordinate along the thickness direction. Derivatives
in other directions do not appear in the equation since any state with
non-zero derivatives in other directions does not correspond to 
minimum $H$. But if the system in fully limited in all directions, last 
equation should have $\bigtriangledown^2$ in place of $d^2/dx^2$. 
In Ref. [12] last equation is solved analytically for
Dirichlet boundary conditions $\Psi(0)=\Psi(1)=0$. The exact solution is
\begin{equation}
\Psi_0(x)={\sqrt{2}k\over \sqrt{1+k^2}}{\rm sn}(2xF(k), k) \ ,
\end{equation}

\noindent in which $k$ is determined by $l$ through
$l=2\sqrt{1+k^2} F(k)$. Here, $F(k)$ is the first kind of complete elliptic
integral, ${\rm sn}(x,k)$ is elliptic sine function. Unfortunately, no simple
compact solution is found yet for other boundary conditions. One can easily
see that the main obstacle comes from the nonlinear term $\Psi^3$ in the
differential equation of $\Psi$. To find approximate solutions of $\Psi$ for
other boundary conditions, the following method can be used.
First of all, we replace $\Psi^3$ by $\lambda\Psi$ and get a solution 
satisfying the same boundary condition. For Dirichlet boundary
conditions, the solution is
\begin{equation}
\Psi_0=A\sin\pi x\ , \hbox{\hspace*{0.8cm}and      } \lambda=1.0-\pi^2/l^2.
\end{equation}

\noindent The constant $A$ can be determined by requiring the mean square
of the deviation caused by the replacement, i.e., the integral
$\int_0^1 dx (\Psi_0^3-\lambda\Psi_0)^2$, to be minimum. Thus one gets
\begin{equation}
\Psi_0(x)=\sqrt{{4\over 3}\left(1-{\pi^2\over l^2}\right)}\sin\pi x\ .
\end{equation}

\noindent Now one can see that the requirement of minimum deviation
caused by the replacement is equivalent to retaining $\sin\pi x$ term but 
neglecting terms with higher frequency in $\Psi_0^3$. Thus, this approximation
is equivalent to the standard functional variation method. The virtue of this
method is that it can be used more simply and in a step-by-step way.
As discussed in Ref. [12],
the vacuum state $\Psi_0=0$ if the reduced thickness $l$ of the film is
less than $\pi$. The existence of minimum reduced thickness of the film
implies a shift of the critical temperature for the finite system from the
bulk one. The exact solutions and the approximate ones are compared in Fig. 1
for $l/\pi$=1.05, 1.10, 1.15, and 1.20. A very good approximation can be seen.
For larger $l$, the same approximative method can be used further after shift
$\Psi_0=\Psi_0^\prime+\sqrt{4(1-\pi^2/l^2)/3}\sin \pi x$ in Eq. (3).   

For Neumann boundary conditions, $\Psi_0^\prime(0)=\Psi_0^\prime(1)=0$,
the vacuum state can also be approximately obtained. The result is
\begin{equation}
\Psi_0=\left\{\begin{array}{ll}
0\ ,& \hbox{\ \ \ for\ \ \     }\ l\le \pi\cr
\sqrt{4(1-\pi^2/l^2)/3}\cos\pi x\ ,& \hbox{\ \ \ for\ \ \   } \pi<l\le 2\pi\cr
\sqrt{4(1-\pi^2/l^2)/5-3/5}\pm\sqrt{8/5-4(1-\pi^2/l^2)/5}\cos\pi x
\ .\hspace*{0.3cm} & \hbox{\ \ \ for\ \ \ } l>2\pi
\end{array}
\right.
\end{equation}

\noindent The two solutions for $l>2\pi$ can be connected through
 $x\leftrightarrow 1-x$.

Then one can consider mixed boundary conditions $\Psi_0(0)=0,
\Psi_0^\prime(1)=0$.
The first order approximation of the solution for vacuum state is
\begin{equation}
\Psi_0(x)=\sqrt{{4\over 3}\left(1-{\pi^2\over 4l^2}\right)}\sin{\pi 
x\over 2}\ .
\end{equation}

As a final example, we give the vacuum state for periodic boundary condition
$\Psi_0(x)=\Psi_0(1+x)$. The approximate vacuum state is
\begin{equation}
\Psi_0(x)=\left\{
\begin{array}{ll}
0\ ,& \hbox{\ \ \ for\ \ \ }l \le 2\pi\cr
\sqrt{4(1-4\pi^2/l^2)/5}(\cos2\pi x\pm \sin 2\pi x)\ ,
& \hbox{\ \ \ for\ \ \ }4\sqrt{6}\pi/3\ge l>2\pi \cr
\sqrt{3(1-32\pi^2/3l^2)/11}\pm\sqrt{4(
1+4\pi^2/l^2)/11}(\cos2\pi x\pm \sin2\pi x) \ . \hspace*{0.3cm}&
\hbox{\ \ \ for\ \ \ }
l>4\sqrt{6}\pi/3
\end{array}
\right.
\end{equation} 

It should be pointed out that $-\Psi_0$ is also a vacuum state of the
system. Then the fluctuations of the system can be around either $\Psi_0$ or
$-\Psi_0$. This is the copy for finite systems of spontaneous symmetry breaking
in $\phi^4$ model.
With the vacuum state $\Psi_0$, one can shift the order parameter $\Psi=
\Psi^\prime+\Psi_0$, then the Hamiltonian $H$ turns out to be
\begin{equation}
H=H[\Psi_0]+{L^3\phi_0^2\over \xi^2}\int d^3 x
{1\over 2}\left[{1\over l^2}(\bigtriangledown \Psi^\prime)^2-{\Psi^\prime}^2+
3\Psi_0^2{\Psi^\prime}^2
+2\Psi_0{\Psi^\prime}^3+{1\over 2}{\Psi^\prime}^4\right] \ .
\end{equation}

\noindent In this expression, $H[\Psi_0]$ has the same form as $H$
in Eq. (2) with $\Psi_0$ in place of $\Psi$. Now the quadratic part
of fluctuation $\Psi$ is positive definite for $l$ larger than
characteristic length, or for temperature enough below the critical
point. Then one sees that the new Hamiltonian can be safely used to
calculate perturbatively fluctuations at low temperature region for
finite systems. For the sake of easier perturbative calculation,
one can use $\langle \Psi_0^2\rangle$ in place of $\Psi_0^2$, and treat
all other terms, $H_I={L^3\phi^2\over \xi^2}\int d^3x \left[
{3\over 2}(\Psi_0^2-\langle \Psi_0^2\rangle){\Psi^\prime}^2
+\Psi_0{\Psi^\prime}^3
+{1\over 4}{\Psi^\prime}^4\right]$, as small perturbations. 
In the lowest order, ignoring contributions from $H_I$, one can get one-point
and two-point and other correlation functions in terms of $\Psi_0$ and
Green's function $G(x,y)$, the inverse of operator $-\bigtriangledown^2/l^2-1
+3\langle\Psi_0^2\rangle$. For example,
\begin{equation}
\begin{array}{l}
\langle \phi(x)\rangle=L^3\phi_0\Psi_0(x)\\
\langle\phi(x)\phi(y)\rangle=L^6\phi_0^2(\Psi_0(x)\Psi_0(y)+{\xi^2\over
L^3\phi_0^2}G(x,y))
\end{array}
\end{equation}

\noindent For system with Dirichlet boundary conditions and under the condition
that $l$ is a little larger than $\pi$, the Green's function $G(x,y)$ is 
\begin{equation}
G(x;y)=\int {d^2p\over (2\pi)^2}
\sum_{i=1}^\infty
{\sin i\pi x_1 \sin i\pi y_1 e^{i{\bf p}\cdot {\bf (r-r^\prime)}}\over
2(1-\pi^2/l^2)-(1-i^2\pi^2/l^2-{\bf p}^2/l^2)}
\end{equation}

\noindent
Here, $x_1, y_1$ are components of coordinates in finite size direction
of two points $x, y$, ${\bf r}$ and ${\bf r}^\prime$ are vectors in other
directions, ${\bf p}$ is the corresponding momentum. An important
feature of the Green's function is that the translational invariance in the
direction with finite length is violated. This result is natural for finite
system. In this time the Green's function cannot be written as function
of the difference of the coordinates of two points. The physical reason
is simple. The Green's function is the response at $x^\prime$
of the system to a source at $x$. When both $x$ and $x^\prime$ are translated
in the same way, the influence of the boundary response changes. Thus, the net
response to the source also changes. So translational invariance
is surely violated. From Eq. (10) one can easily get the vertices needed. Then
a perturbative calculation can be done readily, which is beyond the scope of
this Letter.  

In summary, we showed the importance of local spontaneous symmetry breaking
for finite system in the calculation of fluctuations in phase transition
in low temperature region for such system. The vacuum states are
approximately given for various boundary conditions for film system
within scalar $\phi^4$ model for phase transition.
With the vacuum state, perturbative calculations can be done safely.

This work was supported in part by the NNSF, the Hubei SF and the SECF in
China.

\vskip 2cm
\begin{center}{\Large Figure Caption}\end{center}

{\bf Fig. 1} Comparison between exact solutions and approximate ones
for Eq. (3) under Dirichlet boundary conditions for $l/\pi$=1.05, 1.10, 1.15,
and 1.20. The solid curves correspond to exact solutions, dotted curves are
drawn according to Eq. (6).
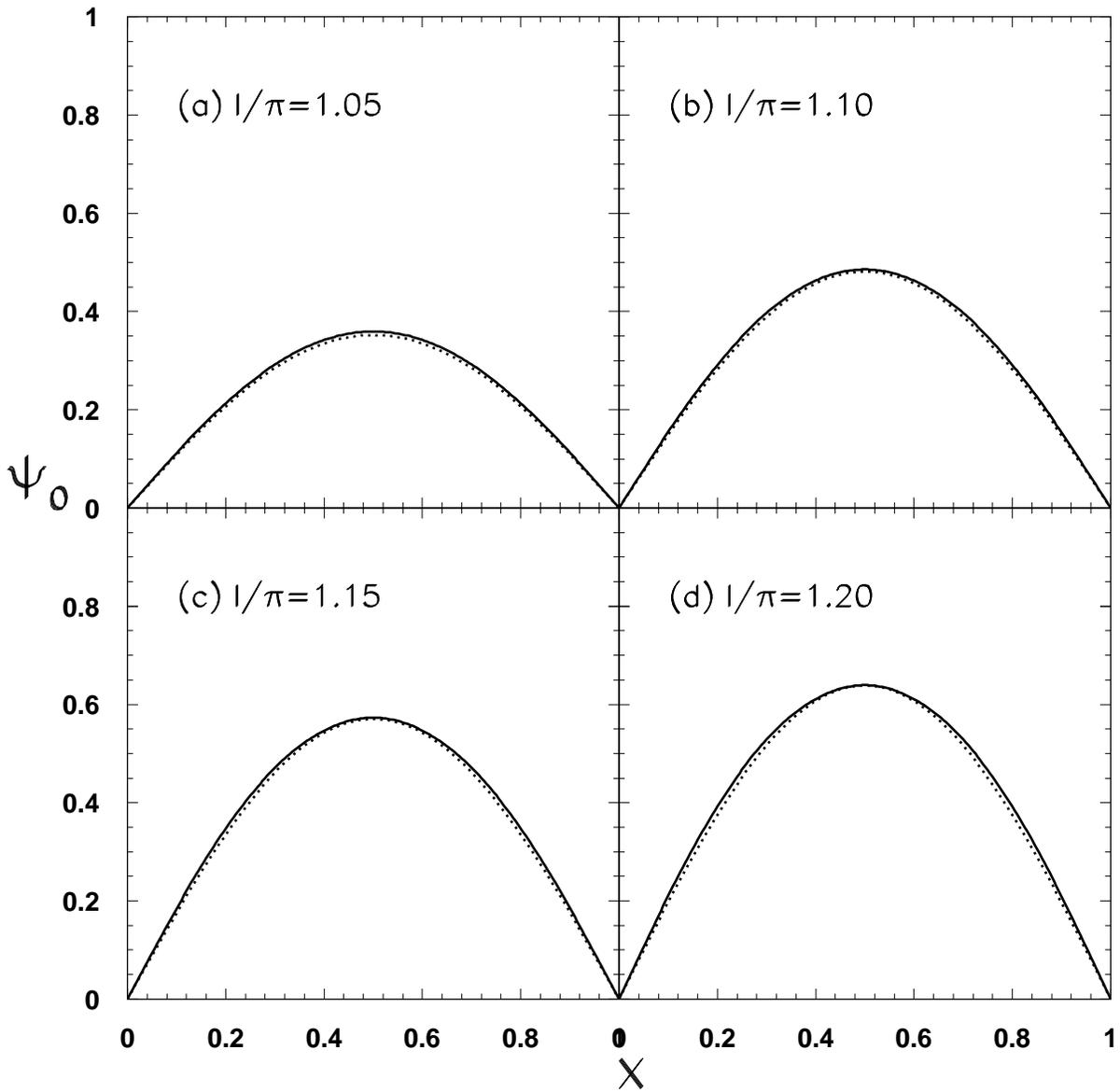
\end{document}